# Enhanced four-wave-mixing with 2D layered graphene oxide films integrated with CMOS compatible micro-ring resonators

*Jiayang Wu, Yunyi Yang, Yang Qu, Linnan Jia, Yuning Zhang, Xingyuan Xu,*

*Sai T. Chu, Brent E. Little, Roberto Morandotti, Baohua Jia,\* and David J. Moss\**

Dr. J. Wu, Dr. Y. Yang, Y. Qu, L. Jia, Y. Zhang, X. Xu, Prof. B. Jia, and Prof. D. J. Moss
Center for Micro-Photonics
Swinburne University of Technology
Hawthorn, VIC 3122, Australia

Dr. Y. Yang, Prof. B. Jia
Adjoint with Centre for Translational Atomaterials
Swinburne University of Technology
Hawthorn, VIC 3122, Australia

Prof. S. T. Chu
Department of Physics, City University of Hong Kong,
83 Tat Chee Avenue, Hong Kong, 999077, SAR, China

Prof. B. E. Little
Xi'an Institute of Optics and Precision Mechanics Precision Mechanics
Chinese Academy of Sciences
Xi'an, 710119, China

Prof. R. Morandotti
INSR-Énergie
Matériaux et Télécommunications
1650 Boulevard Lionel-Boulet, Varennes, Québec, J3X 1S2, Canada

Prof. R. Morandotti
Adjoint with Institute of Fundamental and Frontier Sciences
University of Electronic Science and Technology of China
Chengdu, 610054, China

Abstract

Layered two‑dimensional (2D) graphene oxide (GO) films are integrated with micro-ring resonators (MRRs) to experimentally demonstrate enhanced nonlinear optics in the form of four-wave mixing (FWM). Both uniformly coated and patterned GO films are integrated on CMOS-compatible doped silica MRRs using a large-area, transfer-free, layer-by-layer GO coating method together with photolithography and lift-off processes, yielding precise control of the film thickness, placement, and coating length. The high Kerr nonlinearity and low loss of the GO films combined with the strong light-matter interaction within the MRRs results in a significant improvement in the FWM efficiency in the hybrid MRRs. Detailed FWM measurements are performed at different pump powers and resonant wavelengths for the uniformly coated MRRs with 1−5 layers of GO as well as the patterned devices with 10−50 layers of GO. The experimental results show good agreement with theory, achieving up to ~7.6-dB enhancement in the FWM conversion efficiency (CE) for an MRR uniformly coated with 1 layer of GO and ~10.3-dB for a patterned device with 50 layers of GO. By fitting the measured CE as a function of pump power for devices with different numbers of GO layers, we also extract the dependence of GO's third-order nonlinearity on layer number and pump power, revealing interesting physical insights about the evolution of the layered GO films from 2D monolayers to quasi bulk-like behavior. These results confirm the high nonlinear optical performance of integrated photonic resonators incorporated with 2D layered GO films.

# 1. Introduction

Nonlinear integrated photonic devices offer powerful solutions to generate and process signals all-optically, with far superior processing speed compared to electronic devices, as well as the added benefits of a compact footprint, low power consumption, high stability, and the potential to significantly reduce cost by mass production [1, 2]. Four-wave mixing (FWM), as a fundamental third-order ($\chi^{(3)}$) nonlinear optical process [1, 3], has been widely used for achieving all-optical signal generation and processing, such as wavelength conversion [4, 5], optical frequency comb generation [6, 7], optical sampling [8, 9], quantum entanglement [10, 11], and many others [12-14].

Integrated micro-ring resonators (MRRs), with their strong light confinement in compact micro-scale resonant cavities, are key building blocks for photonic integrated circuits and play important roles in many applications from optical communications to optical interconnects, photonic processing, and biosensing [15-17]. Compared with FWM in integrated waveguides, FWM in integrated MRRs can provide dramatically enhanced conversion efficiencies (CE) due to resonant enhancement of the optical field [18, 19], thus significantly reducing the power requirements. FWM has been demonstrated in integrated MRRs fabricated on III-V platforms including GaAs and AlGaAs [19-21], and also complementary metal-oxide-semiconductor (CMOS) compatible platforms including silicon, silicon nitride, and high index doped silica glass [22-24]. Although silicon has been a leading platform for integrated photonic devices, its strong two-photon absorption (TPA) at near-infrared wavelengths (TPA coefficient: ~0.9 cm/GW [2]) poses a fundamental limitation for FWM in the telecommunications band [1]. CMOS compatible platforms such as silicon nitride and high index doped silica glass have a much weaker TPA (no TPA observed in the telecommunications band even up to extremely high light intensities of ~25 GW/cm$^2$ [2]), although they face limitations in terms of FWM efficiency since their Kerr nonlinearity ($n_2$) is over an order of magnitude smaller than that of silicon [2].

The quest for high-performance nonlinear integrated photonic devices has motivated the use of highly nonlinear materials on chips to overcome the limitations of existing platforms [8, 25]. The giant Kerr nonlinear response of two-dimensional (2D) layered materials such as graphene, graphene oxide (GO), black phosphorus, and transition metal dichalcogenides (TMDCs) has been widely recognized and exploited to implement diverse nonlinear photonic devices with high performance and new capabilities [25-35]. In particular, a 6.8-dB enhancement in the FWM CE was reported for a silicon MRR incorporating a monolayer of doped graphene [31].

Owing to its ease of preparation as well as the tunability of its material properties, GO has become a highly promising member of the graphene family [27, 36-39]. Previously, we reported GO films with a giant Kerr nonlinearity ($n_2$) of about 4 orders of magnitude higher than that of silicon [27, 40], and achieved enhanced FWM CE in GO-coated doped silica waveguides of up to 6.9 dB for a 1.5-cm-long waveguide uniformly coated with 2 layers of GO [29]. Moreover, GO has a material absorption that is over 2 orders of magnitude lower than graphene [29] as well as a large bandgap (2.1−2.4 eV) that yields a low TPA in the telecommunications band [41, 42]. Recently, we achieved highly precise control of the placement, thickness, and length of the GO films coated on integrated photonic devices by using a large-area, transfer-free, layer-by-layer GO coating method together with standard photolithography and lift-off processes [43]. This overcomes critical fabrication limitations in terms of layer transfer and precise patterning for on-chip integration of 2D materials and represents a significant advance towards manufacturing integrated photonic devices incorporated with 2D materials.

In this paper, we use our GO fabrication techniques to demonstrate enhanced FWM in MRRs integrated with 2D layered GO films. Owing to the strong light-matter interaction in the MRRs incorporating highly nonlinear GO films, the FWM efficiency in the hybrid MRRs is significantly improved. We perform FWM measurements at different pump powers and resonant wavelengths for CMOS-compatible doped silica MRRs uniformly coated with 1−5 layers of GO as well as devices patterned with 10−50 layers of GO, achieving up to ~7.6-dB enhancement in the FWM CE for an MRR uniformly coated with 1 layer of GO and ~10.3-dB for a device patterned with 50 layers of GO. Compared with the uniformly coated MRRs, the MRRs with patterned GO provide more flexibility to balance the trade-off between FWM enhancement and linear loss, which is critical for optimizing the CE and fully exploiting the high Kerr nonlinearity of 2D GO films. We also fit the measured CE based on FWM theory and obtain the dependence of the Kerr nonlinearity of the GO films on the number of layers and on the pump power. Our results reveal physical insights and trends of the layered GO films in evolving from 2D monolayers to quasi bulk-like behavior. These results confirm the effectiveness of introducing 2D layered GO films into integrated photonic resonators to improve the performance of nonlinear optical processes.

## 2. Device fabrication and characterization

**Figure 1(a)** shows a schematic of an integrated MRR incorporating a GO film. The MRR was fabricated on a high index doped silica glass platform using CMOS compatible fabrication processes [18, 44] with chemical mechanical polishing (CMP) used as the last step to remove the upper cladding, so as to enable GO film coating on the top surface of the MRR. Benefiting from extraordinarily low linear and nonlinear loss, high index doped silica glass has been a successful integrated platform for nonlinear photonic devices [6, 10, 11, 45, 46]. The $n_2$ of the high index doped silica glass (~$1.3 \times 10^{-19}$ m$^2$/W) is lower than that of silicon (~$4.5 \times 10^{-18}$ m$^2$/W), while its negligible nonlinear loss even up to extremely high light intensities yields a nonlinear figure of merit (>>1) that is much higher than that of silicon (~0.3) [2]. The CMOS compatible fabrication processes also make our devices comparable, in terms of fabrication maturity, to those implemented in silicon optoelectronic devices [2, 18].

The coating of 2D layered GO films was achieved via a solution-based method that yields layer-by-layer GO film deposition on a dielectric substrate, as reported previously [29, 41]. Four steps for the in-situ assembly of monolayer GO films were repeated to construct multilayer films on a target substrate. **Figure 1(b)** shows a scanning electron microscope (SEM) image of a 2D layered GO film on a silica substrate, with up to 5 layers of GO. Our GO coating approach, unlike the sophisticated transfer processes employed for coating other 2D materials such as graphene [47, 48], enables large-area, transfer-free, and high-quality GO film coating on integrated photonic devices, with highly scalable fabrication processes and precise control of the number of GO layers (i.e., GO film thickness).

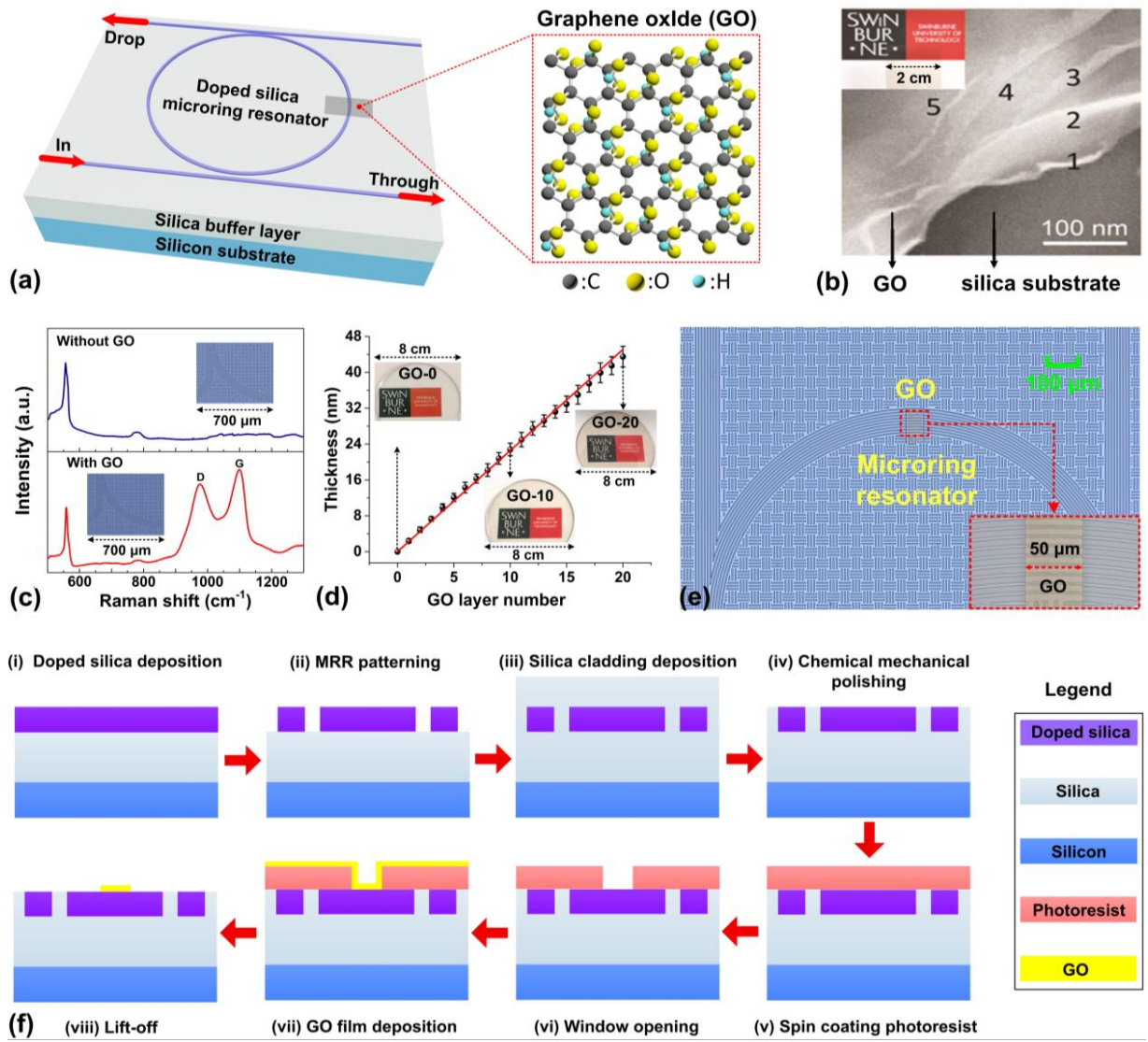

**Figure 1.** (a) Schematic illustration of GO-coated integrated MRR. Inset shows schematic atomic structure of GO. (b) Scanning electron microscope (SEM) image of 2D layered GO film. The numbers 1−5 refer to the number of GO layers for that part of the image. Inset shows 5 layers of GO on a silica substrate. (c) Raman spectra of an integrated chip without GO and with 2 layers of GO. Insets show the corresponding microscope images. (d) Measured GO film thickness versus GO layer number. Insets show the images of a silica circular substrate uniformly coated with 0 (uncoated), 10, and 20 layers of GO. (e) Microscopic image of an integrated MRR

patterned with 50 layers of GO. Inset shows zoom-in view of the patterned GO film. (f) Schematic illustration showing the fabrication process flow for an integrated MRR with patterned GO film.

**Figure 1(c)** shows the measured Raman spectra of an integrated chip (including doped silica MRRs) without GO and with 2 layers of uniformly coated GO film. The presence of the representative D (1345 cm$^{-1}$) and G (1590 cm$^{-1}$) peaks of GO confirms the integration of GO film onto the top surface. **Figure 1(d)** shows the thickness of GO films versus the number of layers, characterized by atomic force microscopy (AFM). The plots show the average of measurements on three samples and the error bars reflect the variations. The GO film thickness shows a nearly linear relationship with the layer number, with a thickness of ~2.25 nm on average for each layer. The insets present images of a silica substrate uniformly coated with 0 (uncoated), 10, and 20 layers of GO, showing large-area GO film coating with high uniformity.

Using this GO coating method, we achieved GO patterning on integrated photonic devices via photolithography and lift-off processes [43]. **Figure 1(e)** shows microscopic images of an integrated MRR patterned with 50 layers of GO (~50 µm pattern length). Note that only the center ring of the 9 concentric rings (see inset) was coupled to through/drop bus waveguides to form an MRR − the rest were to aid in visual identification. We achieved GO film patterning with lengths down to ~150 nm via e-beam lithography and accurate placement on the MRR (deviation < 20 nm) by using gold alignment markers [43]. The schematic illustration of the fabrication process flow is provided in **Figure 1(f)**. The combination of GO coating with photolithography and lift-off allows precise control of the film placement, size, and thickness on integrated devices. This, in turn, allowed for investigating the layer dependent material properties of 2D layered GO films as well as optimizing the device performance such as FWM CE. Finally, along with large-area uniform coating capability, our fabrication techniques enable large-scale integrated devices incorporated with GO films. This is not only for nonlinear photonic devices but also for integrated devices based on other extraordinary material properties of GO such as broadband photoluminescence / light absorption [37, 39], strong material anisotropy [43], and flexible bandgap engineering [27, 42, 49].

We fabricated and tested two types of GO-coated MRRs − uniformly coated with 1−5 layers of GO and patterned with 10−50 layers of GO (~50 μm pattern length). The radius of the MRR was ~592 μm, corresponding to a free spectral range (FSR) of ~0.4 nm (~49 GHz) for the uncoated MRR. Compared with the uniformly coated MRRs, the patterned MRRs enabled us to test the device performance with shorter GO film lengths but higher film thicknesses (up to 50 layers). **Figures 2(a)** and **(b)** show the measured transmission spectra of the MRR with uniformly coated and patterned GO films, all integrated with the same doped silica MRR and measured using a low-power (0 dBm) transverse electric (TE) polarized continuous-wave (CW) light. The coated GO films could be easily removed by plasma oxidation, and so the same uncoated MRR was reused for coating GO films with different numbers of GO layers. We used a 16-channel single-mode fiber (SMF) array to butt couple the CW light near 1550 nm into and out of the MRR chip. The mode coupling loss between the SMF array and the doped silica waveguide was ~8 dB/facet, which can readily be reduced to < 2.0 dB/facet with on-chip mode convertors [6, 50]. For the uniformly coated devices, the GO film thicknesses (1−5 layers, i.e., ~2−10 nm) were very small as compared with the geometry of the doped silica waveguide (~2 μm width ×1.5 μm height) and so their influence on the MRR's coupling condition was negligible. The difference between the coupling strength of the uncoated MRR and the uniformly coated MRR with 5 layers of GO was ~0.3%. For the patterned devices, on the other hand, the GO films were located outside the coupling area, and so they did not affect the MRR's coupling condition. We chose the TE polarization for the experiments because it supported the in-plane interaction between the evanescent field and the thin GO film, which is much stronger than the out-of-plane interaction due to the large optical anisotropy of 2D layered materials [43, 51].

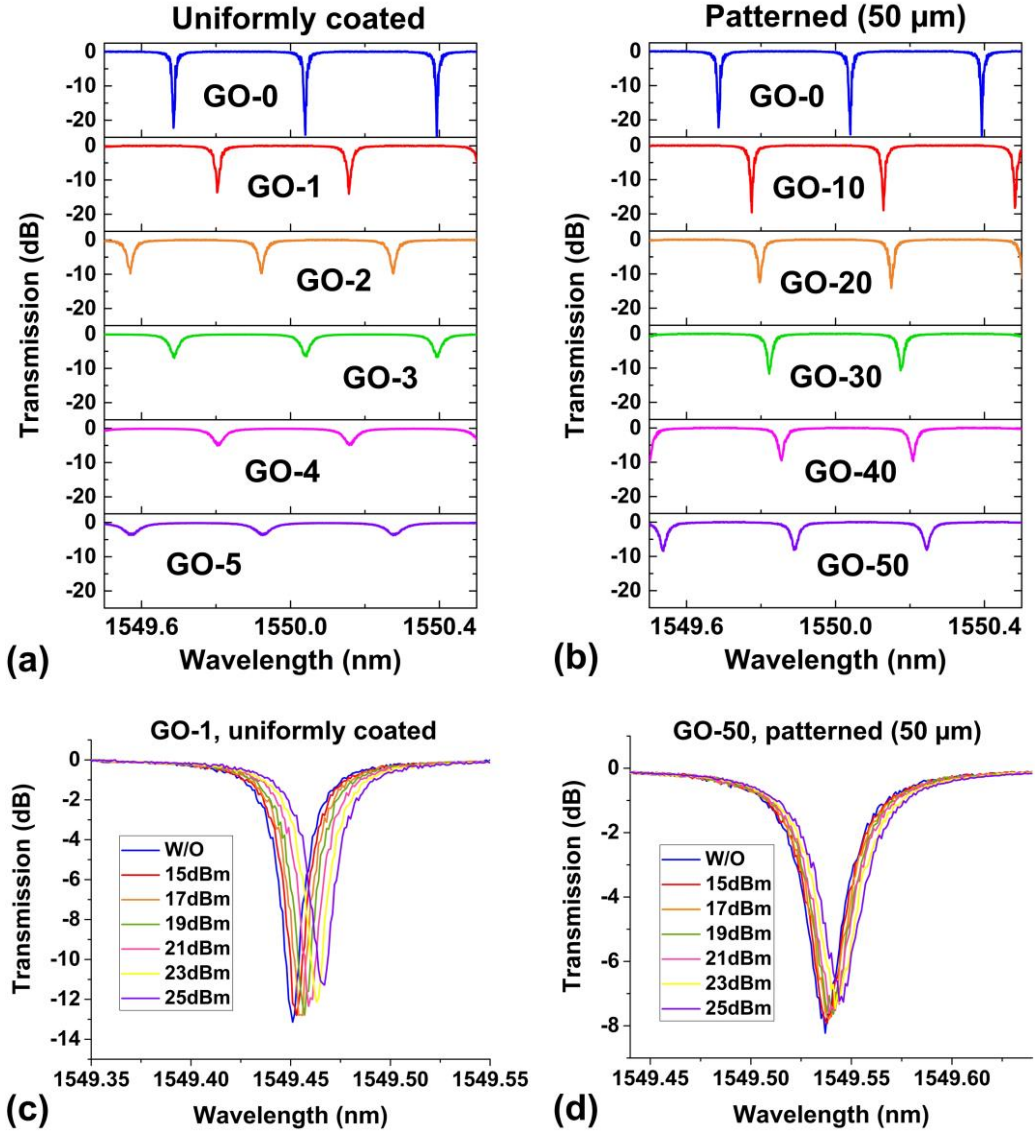

**Figure 2.** (a)−(b) Transmission spectra of an integrated MRR uniformly coated with 1−5 layers of GO and patterned with 10−50 layers of GO measured using a low-power CW light, respectively. The measured transmission spectrum of the uncoated MRR (GO-0) is also shown for comparison. (c)−(d) Transmission spectra of the MRRs with 1 layer of uniformly coated and 50 layers of patterned GO measured using a low-power CW probe when another high-power CW pump was injected into a resonance around 1550.18 nm, respectively. The values of 15−25 dBm represent the incident pump powers. The transmission spectra measured using the low-power CW probe without the high-power CW pump (W/O) are also shown for comparison.

We also measured the CW power-dependent transmission spectra of the GO-coated MRRs up to high powers (25 dBm) using a pump-probe method where we injected a high-power CW pump into a resonance around 1550.18 nm and a low-power probe (0 dBm, also used for measuring the spectra in **Figures 2(a)** and **(b)**) to scan the spectra around another resonance. **Figures 2(c)** and **(d)** show the transmission spectra of the MRRs with 1 layer of uniformly coated and 50 layers of patterned GO films for different incident pump powers (15−25 dBm, excluding mode coupling loss between the SMF array and the input bus waveguide as well as GO-induced propagation loss of the input bus waveguide for the uniformly coated devices). There were small but observable changes in the resonance wavelength and notch depth with pump power for the coated MRRs, reflecting a change in the GO material properties (e.g., refractive index and loss). For the uncoated MRR, on the other hand, we could not observe any obvious changes. The changes we observed were not permanent – the transmission spectra recovered to those in **Figures 2(a)** and **(b)** when we turned off the high-power CW pump, with the spectra in **Figures 2(c)** and **(d)** being repeatable (see Section 4 for detailed discussion).

**Figures 3(a)** and **(c)** show the extinction ratios (ERs) and quality factors (Qs) extracted from the measured power-dependent transmission spectra of the MRRs uniformly coated with 1−5 layers of GO, respectively, along with the results for the uncoated MRR. The extracted power-dependent ERs and Qs of the MRRs patterned with 10−50 layers of GO are provided in **Figures 3(b)** and **(d)**, respectively. The plots show the average of 5 resonances around 1548.5 nm and the error bars reflect the variations. The uncoated MRR had high ERs (> 20 dB) and relatively high Qs (~70,000) (although significantly less than those of doped silica MRRs with silica cladding [18, 24]). As expected, both the ER and Q decreased with the number of GO layers, reflecting the additional loss induced by the GO film. As the input CW power was increased, the MRRs with both uniformly coated and patterned GO films exhibited a clear decrease in ER and Q, whereas the uncoated MRR did not. The propagation loss of the GO hybrid waveguides was obtained by using the scattering matrix method [52, 53] to fit the ERs and Qs in **Figures 3(a)−(d)**, and is shown for 1−5 layers and 10−50 layers in **Figures 3(e)** and **(f)**, respectively. Note that the propagation loss obtained from the GO-coated MRRs is different from that obtained from the GO-coated waveguides [29, 43] when considering the power-dependent loss changes of the GO films, since the light intensity in the MRRs is significantly higher due to resonant enhancements.

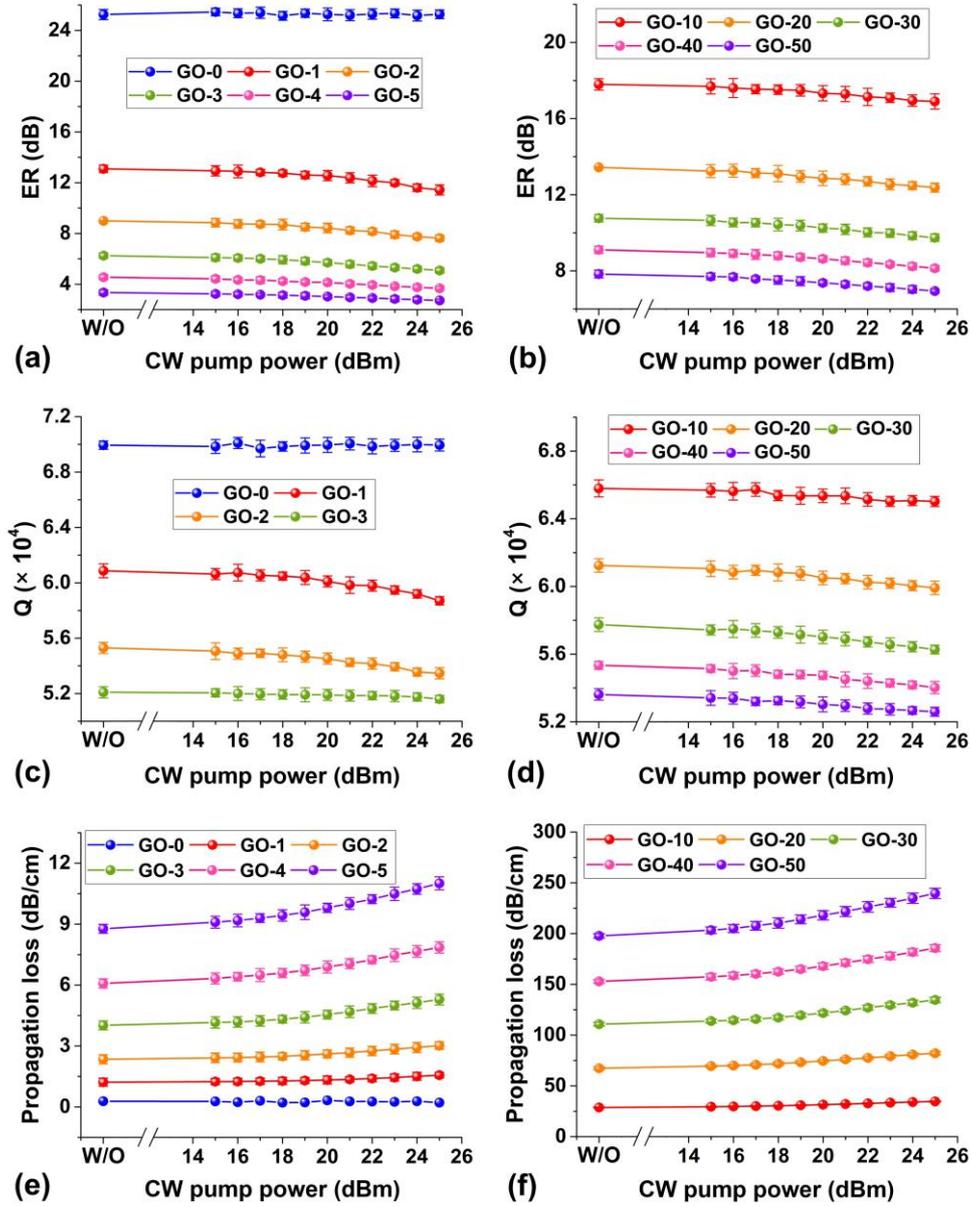

**Figure 3.** Power dependent (a)−(b) extinction ratios (ERs) and (c)−(d) quality factors (Qs) for the MRRs with 1−5 layers of uniformly coated and 10−50 layers of patterned GO films, respectively. The ERs and Qs of the uncoated MRR (GO-0) are also shown for comparison. The Qs are not shown when the ERs are < 5 dB. (e)−(f) Fit propagation loss obtained from (a)−(d), respectively. The values of 15−25 dBm represent the incident pump powers. The data points for W/O correspond to the values measured using a low-power CW probe without the high-power CW pump.

The low-power (W/O in **Figures 3(e)**) propagation loss of the uncoated waveguide and the waveguide with a monolayer of GO was ~0.26 dB/cm and ~1.27 dB/cm, respectively, corresponding to an excess propagation loss of ~1 dB/cm induced by the GO film. This is over 2 orders of magnitude lower than that of integrated waveguides coated with graphene [54, 55], indicating the low material absorption of GO and its strong potential for the realization of high-performance nonlinear photonic devices. The propagation loss increased with the GO layer number − a combined result of mode overlap and several other possible effects such as increased scattering loss and absorption induced by imperfect contact between the multiple GO layers as well as interaction between the GO layers, as reported previously [43]. As the input CW power was increased, all the MRRs showed an increased propagation loss except for the uncoated MRR, which further confirms the change of GO material properties with light power in the GO-coated MRRs.

## 3. FWM experiment

**Figure 4** shows the experimental setup used to measure FWM in the MRRs. Two CW tunable lasers separately amplified by erbium-doped fiber amplifiers (EDFAs) were used as the pump and signal sources, respectively. In each path, there was a polarization controller (PC) to ensure that the input light was TE-polarized. The pump and signals were combined with a 3-dB fiber coupler before being coupled into the MRR, which was mounted on a temperature control stage (TCS) to avoid thermal resonance drift and to maintain the wavelength alignment of the resonances to the CW pump and signal. An optical isolator was employed to prevent the reflected light from damaging the laser source. The signal output from the drop port of the MRR was sent to an optical spectrum analyzer (OSA) with a variable optical attenuator (VOA) being inserted before the OSA to prevent high-power damage.

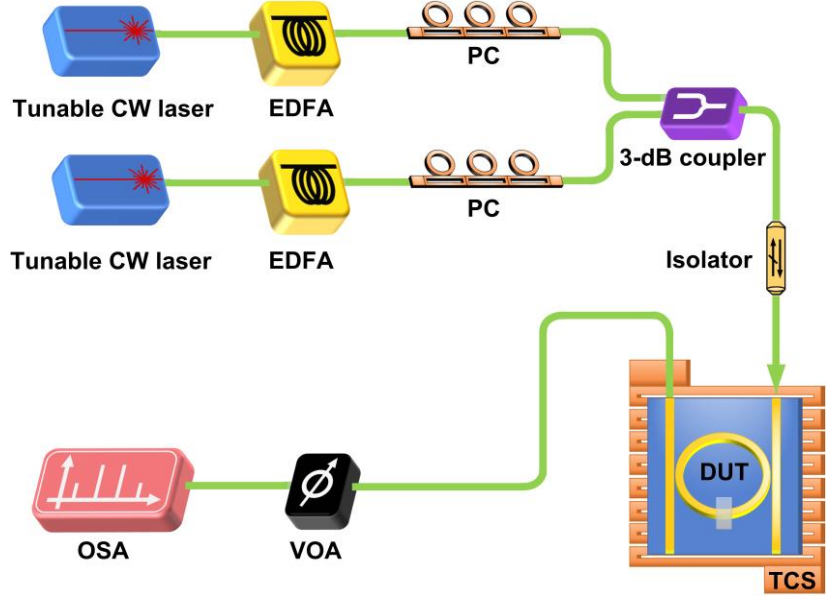

**Figure 4.** Experimental setup for FWM measurements in integrated MRRs. EDFA: erbium-doped fiber amplifier. PC: polarization controller. DUT: device under test. TCS: temperature controller stage. VOA: variable optical attenuator. OSA: optical spectrum analyzer.

**Figure 5(a)** shows the FWM spectra of the MRRs uniformly coated with 1−5 layers of GO, together with the FWM spectrum of the uncoated MRR. For comparison, we kept the same pump power of ~22 dBm coupled into the MRRs after excluding the mode coupling loss between the SMF array and the input bus waveguide as well as the GO-induced propagation loss of the input bus waveguide. The pump and signal had the same power and were separated in wavelength by 2 FSRs of the MRRs. As compared with the uncoated MRR, the GO-coated MRRs had an additional insertion loss (defined as the excess insertion loss of the GO-coated MRRs over the uncoated MRR), while the MRRs with 1 and 2 layers of GO clearly show enhanced idler output powers. The CE (defined as the ratio of the output power of the idler to the input power of the signal, i.e., $P_{idler,\,out}/P_{signal,\,in}$) of the MRR without GO and with 1 layer of GO were ~-48.4 dB and ~-40.8 dB, respectively, corresponding to a CE enhancement of 7.6 dB for the GO-coated MRR. **Figure 5(b)** shows the FWM spectra of the MRRs with 10−50 layers of patterned GO. The GO coating length was ~50 μm and the input pump power (22 dBm) was the same as that in **Figure 5(a)**. The MRRs with patterned GO films also had an additional insertion loss as compared with the uncoated MRR, and the results for all the tested

GO layer numbers show enhanced idler output powers. In particular, there is a maximum CE enhancement of ~10.3 dB for the MRR patterned with 50 layers of GO.

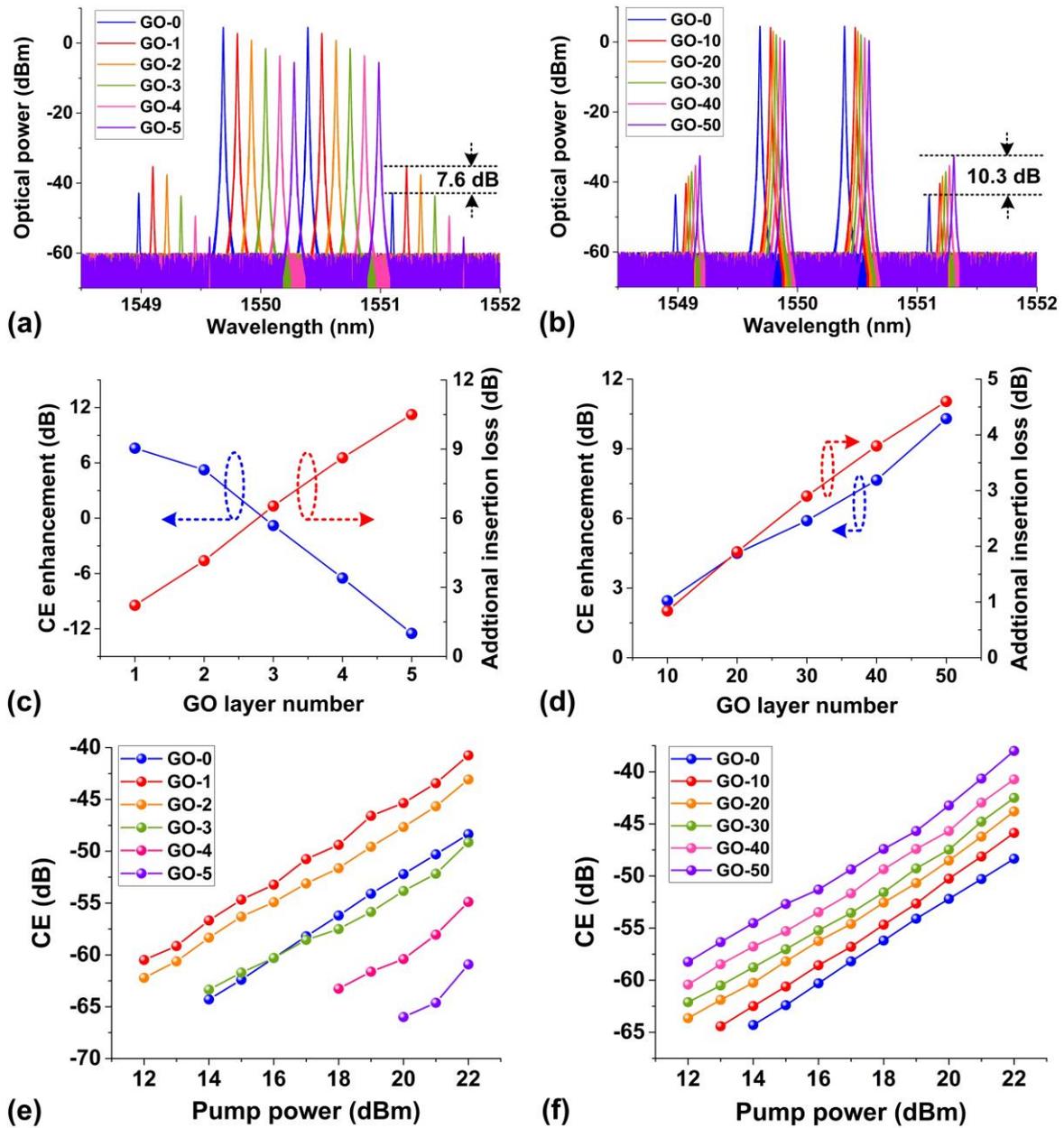

**Figure 5.** (a)−(b) Optical spectra of FWM at a pump power of 22 dBm for the MRRs with 1−5 layers of uniformly coated and 10−50 layers of patterned GO films, respectively. (c)−(d) CE enhancement and additional insertion loss extracted from (a) and (b), respectively. (e)−(f) CE versus pump power for the MRRs with 1−5 layers of uniformly coated and 10−50 layers of patterned GO films, respectively. In (a), (b), (e), and (f), the results for uncoated MRR (GO-0) are also shown for comparison.

The CE enhancement and additional insertion loss extracted from **Figures 5(a)** and **(b)** are shown in **Figures 5(c)** and **(d)**, respectively. As can be seen, the additional insertion loss increases with the GO layer number, showing agreement with the increase of propagation loss with GO layer number in **Figures 3(e)** and **(f)**. For the MRRs with uniformly coated GO, the CE enhancement decreases with the GO layer number, whereas the MRRs with patterned GO shows the opposite trend. This could reflect the trade-off between FWM enhancement (which dominates for the patterned MRRs with a short GO coating length) and loss (which dominates for the uniformly coated MRRs with a much longer GO coating length) in the GO-coated MRRs (see Section 4 for detailed discussion).

**Figures 5(e)** and **(f)** show the measured CE versus pump power for the MRRs with uniformly coated and patterned GO films, respectively. We could not measure a CE lower than -66 dB since the generated idler was below the noise floor. For the uncoated MRR, the dependence of CE versus pump power shows a nearly linear relationship. In contrast, for the GO-coated MRRs, the measured CE shows a relatively obvious deviation from the linear relationship with pump power. Similar to the power dependent loss in **Figures 3(e)** and **(f)**, this is also a reflection of the change in GO material properties with light power (see Section 4 for detailed discussion). As the pump power increased, the measured CE increased with no obvious saturation for the uncoated MRR and the GO-coated MRRs, indicating the low TPA of both the high index doped silica glass and the GO films. Unlike graphene that has a metallic behavior and a zero bandgap, GO is a dielectric that has a large bandgap of 2.1−2.4 eV [41, 42], which results in low linear and nonlinear light absorption in spectral regions below the bandgap, in particular, featuring greatly reduced TPA in the telecommunications band. This represents another important advantage of GO for implementing high-performance nonlinear photonic devices. In theory, the GO film with a bandgap > 2 eV should have no absorption for light at a wavelength near 1550 nm (~0.8 eV). We therefore infer that the loss of the coated GO films is mainly induced by light absorption from localized defects as well as scattering loss stemming from film unevenness and imperfect contact between the multilayers [29, 43]. The relatively low FWM CE is mainly induced by the low Kerr nonlinearity of doped silica. Although our GO-coated MRRs were based on a CMOS compatible doped silica platform, these GO films can readily be introduced into other integrated platforms (e.g., silicon and silicon nitride) offering significantly enhanced mode overlap between GO and waveguide with reduced waveguide dimensions.

**Figures 6(a)−(c)** show the FWM spectra versus Δλ (wavelength spacing between pump and signal) for the uncoated MRR, the uniformly coated MRR with 1 layer of GO, and the patterned MRR with 50 layers of GO, respectively. The pump power was 22 dBm with the wavelength tuned to a resonance near 1550 nm and the signal wavelength detuned from 2 to 20 FSRs. The measured CE versus Δλ is depicted in **Figure 6(d)** where we see that, for all three MRRs, the CE only shows a slight decrease with Δλ (< 1.6 dB for Δλ / FSR = 20), reflecting the low dispersion of the doped silica MRR (~2 orders of magnitude lower than silicon MRRs [24]) and the GO-coated MRRs, thus enabling effective phase matching for broadband FWM.

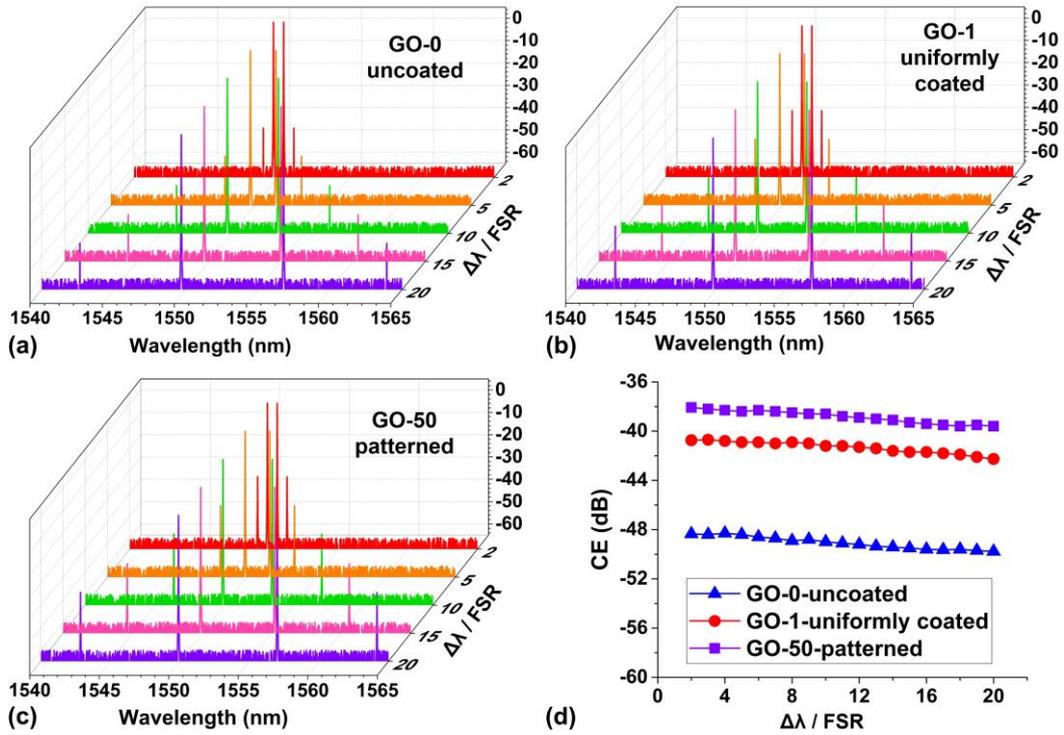

**Figure 6.** (a)−(c) Optical spectra of FWM at different resonant wavelengths for the uncoated MRR, the MRR uniformly coated with 1 layer of GO, and the MRR patterned with 50 layers of GO, respectively. Δλ and FSR represent the wavelength spacing between pump and signal and the free spectral ranges of the MRRs, respectively. (d) Measured CE versus Δλ/FSR for the MRRs in (a)−(c). The pump power in (a)−(d) was 22 dBm.

## 4. Theoretical analysis and discussion

We used the theory from Ref. [19] to model FWM in the hybrid integrated MRRs. Assuming negligible depletion of the pump and signal powers due to the generation of the idler, the FWM CE of the GO-coated MRR is given by

$$CE_{MRR} = \frac{P_{idler,\,out}}{P_{signal,\,in}} = CE_{WG} \cdot FE_p^4 \cdot FE_s^2 \cdot FE_i^2, \qquad (1)$$

where $P_{idler,\,out}$ and $P_{signal,\,in}$ are the output power of the idler and input power of the signal, respectively. $CE_{WG}$ is the FWM CE in a straight waveguide with the same length as the circumference of the MRR, which can be expressed as

$$CE_{WG} = |A_i(L)|^2 / |A_s(0)|^2, \qquad (2)$$

where $A_i(L)$ and $A_s(0)$ are the amplitudes of the idler and signal waves along the light propagation direction, respectively, and $L$ is the waveguide length (i.e., circumference of the MRR). For the waveguide with uniformly coated GO, $CE_{WG}$ was calculated by solving coupled differential equations for FWM processes based on the theory in Ref. [29]. For the waveguide with patterned GO, $CE_{WG}$ was calculated by dividing the waveguide into coated and uncoated parts and solving the FWM differential equations in each part. In our simulation, the uncoated doped silica waveguide has an anomalous dispersion (group velocity dispersion $\beta_2 \approx -3.9 \times 10^{-26}$ s$^2$/m), and the hybrid waveguides with GO films show slightly enhanced anomalous dispersion (e.g., $\beta_2 \approx -3.92 \times 10^{-26}$ s$^2$/m for the hybrid waveguide with 5 layers of GO) for better phase matching. It is also worth mentioning that the effect of phase matching is negligible in our case since $L\beta_2\Delta\omega^2 \ll 1$ [28, 29], where $L$ is the circumference of the MRR, and $\Delta\omega < 1.6 \times 10^{13}$ rad/s (i.e., ~20 nm wavelength spacing) is the angular frequency detuning range.

$FE_{p,s,i}$ in Eq. (1) are resonant field enhancement factors for the pump, signal, and idler, respectively, which can be written as

$$FE_{p,\,s,\,i} = \kappa_r \cdot t_r / [1 - t_r^2 \cdot A_r \cdot \exp(j \cdot \phi_{p,\,s,\,i})], \qquad (3)$$

where $\kappa_r$ and $t_r$ are the field coupling and transmission coefficients between the MRR and the bus waveguides, respectively. $A_r$ is the round-trip field transmission factor and $\phi_{p,\,s,\,i}$ are the round-trip phase shift of the pump, signal, and idler, respectively, which can be further expressed as

$$A_r = \begin{cases} \exp\left(-\frac{1}{2}\alpha_c L\right), & \text{for uniformly coated MRR} \\ \exp\left(-\frac{1}{2}\alpha_u L_u\right)\exp\left(-\frac{1}{2}\alpha_c L_c\right), & \text{for patterned MRR} \end{cases} \quad (4)$$

$$\phi_{p,s,i} = \begin{cases} k_{pc,sc,ic} L, & \text{for uniformly coated MRR} \\ k_{pu,su,iu} L_u + k_{pc,sc,ic} L_c, & \text{for patterned MRR} \end{cases} \quad (5)$$

where $\alpha_{c,u}$ and $L_{c,u}$ are the loss factors (including both linear and nonlinear loss) and lengths of the coated and uncoated waveguide segments, respectively, and $k_{pc,sc,ic}$ and $k_{pu,su,iu}$ are the wavenumbers of the pump, signal, and idler for the coated and uncoated segments, respectively. For $\alpha_c$, we used the power dependent loss for each layer number obtained from **Figures 3(e)** and **(f)**. Note that 15−25 dBm of power for a single CW pump in **Figures 3(e)** and **(f)** corresponds to 12−22 dBm of pump power in **Figures 5(e)** and **(f)** since we used the same power for the pump and signal in the FWM measurements and so there were the same overall CW powers in the MRRs (the idler power could be neglected). Eq. (3) is equivalent to that reported in our previous work [18] and is slightly different from that in Refs. [19, 31] due to the use of a 4-port MRR in our case.

Based on Eqs. (1)−(5), we fit the experimentally measured CE in **Figures 5(e)** and **(f)** to obtain the nonlinear parameter $\gamma$ of the GO hybrid waveguide. We then obtained the Kerr coefficient $n_2$ of the layered GO film using the following equation [29]

$$\gamma = \frac{2\pi}{\lambda} \frac{\iint_D n_0^2(x,y) n_2(x,y) S_z^2 dxdy}{\left[\iint_D n_0(x,y) S_z dxdy\right]^2}, \quad (6)$$

where $\lambda$ is the pump wavelength, $D$ is the integral domain over the material regions with the fields, and $S_z$ is the time-averaged Poynting vector calculated using Lumerical FDTD commercial mode solving software. $n_0(x, y)$ and $n_2(x, y)$ are the linear refractive index and $n_2$ profiles over the waveguide cross section, respectively. In our FDTD simulation, we used the material refractive index of doped silica and GO measured via spectral ellipsometry and neglected the small modal distribution difference induced by the slight change of GO's refractive index with power during FWM. This work was performed in the regime close to degeneracy where the three FWM frequencies (pump, signal, idler) were close together (compared with any variation in $n_2$ arising from its dispersion [29]). We therefore used $n_2$ instead of $\chi^{(3)}$ in our subsequent analysis and discussion of the material third-order nonlinearity. The values of $n_2$ for pure silica and doped silica used in our calculation were $2.60 \times 10^{-20}$ m$^2$/W [2] and $1.28 \times 10^{-19}$ m$^2$/W, respectively, the latter obtained by fitting the experimental results from the uncoated MRR. Note that $\gamma$ in Eq. (6) is an effective nonlinear parameter weighted by not only $n_2(x, y)$ but also $n_0(x, y)$ in different material regions, which is more accurate for high-index-contrast hybrid waveguides studied here as compared with the theory in Ref. [31]. The FWM CE can be further optimized by redesigning the waveguide cross section to improve mode overlap with the GO film as well as changing the coupling strength to achieve a better field enhancement.

**Table I. FWM performance comparison for integrated photonic devices incorporating graphene and GO. Si: silicon. PhC: photonic crystal. WG: waveguide. MRR: microring resonator.**

| 2D Material | Integrated platform & device | 2D film thickness & layer No. | Loss induced by monolayer 2D material [a] | $n_2$ of 2D material ($\times n_2$ of Si) [b] | Max. CE enhancement (dB) | Ref. |
|---|---|---|---|---|---|---|
| Graphene | Si PhC cavity | ~1 nm 1 layer | ~500 dB/cm | ~2 x 10$^4$ | >20 | [25] |
| Graphene | Si PhC WG | ~1 nm 1 layer | ~500 dB/cm | ~2 x 10$^4$ | ~8 | [56] |
| Graphene | Si WG | ~0.5 nm 1 layer | ~300 dB/cm | ~2 x 10$^5$ | ~4.8 | [57] |
| Graphene | Si MRR | 1 layer | N/A [c] | ~3 x 10$^4$ | ~6.8 | [31] |
| GO | Doped silica WG | ~4 nm 2 layers | ~1 dB/cm | ~3 x 10$^3$ | ~6.9 | [29] |
| GO | Doped silica MRR | ~2−100 nm 1−50 layers | ~1 dB/cm | ~3 x 10$^3$ −5 x 10$^3$ | ~10.3 | This work |

a) The loss is the linear propagation loss of the hybrid WG after excluding the propagation loss of the bare WG. b) $n_2$ of silicon = $4.5\times10^{-18}$ m$^2$/W (Ref. [2]). c) The information is not provided.

**Table I** compares the FWM performance of integrated photonic devices incorporating graphene and GO. Here, we focus on integrated devices, although we note that strong FWM has also been demonstrated for optical fibers and spatial-light systems incorporating 2D materials [58-60]. As can be seen, although the $n_2$ of GO is about 1 order of magnitude lower than graphene, the loss induced by GO is over two orders of magnitude lower. Further, the $n_2$ of GO is still over 3 orders of magnitude higher than silicon, suggesting that the GO films can also be introduced into other integrated platforms (e.g., silicon and silicon nitride) to improve the FWM performance.

**Figure 7(a)** shows $n_2$ of GO films versus layer number obtained from the measured CEs in **Figures 5(e)** and **(f)** for fixed pump powers of 12 dBm and 22 dBm. The values of $n_2$ are about 4 orders of magnitude higher than that of silicon and agree reasonably well with our previous waveguide FWM experiments [29] and Z-scan measurments [40]. **Table II** compares the measured $n_2$ of GO in our work with previous reports on solid GO films. Note that our work is the first study of the dependence of $n_2$ on the number of layers for 2D GO films, which is challenging for Z-scan measurements because of the weak response of extremely thin 2D films [27, 40]. We also note that the Kerr nonlinearity of GO and GO nanocomposite has been studied in Refs. [61-63] via the Z-scan method, with the measured $n_2$ being lower than those in **Table II**, possibly due to their samples being dispersed in solutions in contrast to solid films in our case. The high $n_2$ of GO films highlights their strong Kerr nonlinearity for not only FWM but also other third-order ($\chi^{(3)}$) nonlinear processes such as third harmonic generation, self-phase modulation, and cross-phase modulation [12, 64-66].

Table II. Comparison of $n_2$ of GO film. WG: waveguide. MRR: microring resonator.

| Material | Wavelength [nm] | Film thickness | $n_2$ (×10$^{-14}$ m$^2$/W) | Method | Ref. |
|---|---|---|---|---|---|
| GO | ~800 | ~2 μm | ~70 | Z-scan | [27] |
| EGO[a] | ~800 | ~300 nm | ~5.7-36.3 | Z-scan | [67] |
| GO | ~1550 | ~1 μm | ~4.5 | Z-scan | [40] |
| GO | ~1550 | ~4 nm | ~1.5 | FWM in WG | [29] |
| GO | ~1550 | ~2−100 nm | ~1.2−2.7 | FWM in MRR | This work |

a) EGO: electro-chemically derived graphene oxide

In **Figure 7(a)**, some $n_2$ values at 12 dBm are not shown since the generated idlers were below the noise floor. $n_2$ (both at 12 dBm and 22 dBm) decreases with GO layer number, showing a similar trend to $WS_2$ measured by a spatial-light system [68]. This is probably due to increased inhomogeneous defects within the GO layers as well as imperfect contact between the multiple GO layers. We also note that the decrease in $n_2$ with GO layer number becomes more gradual for thicker GO films. This could reflect the transition of the GO film properties towards the bulk material (with a thickness independent $n_2$) for thick films. At 22 dBm, $n_2$ is higher than at 12 dBm, indicating a more significant change of the GO optical properties with inceasing power. The rate of decrease in $n_2$ with GO layer number at 22 dBm is lower than that at 12 dBm, possibly reflecting the fact that thicker GO films are more easily affected by light power.

**Figures 7(b)** and **(c)** show $n_2$ of GO films versus pump power obtained from the measured CEs in **Figures 5(e)** and **(f)**, indicating that, in contrast with the monotonic decrease in $n_2$ with GO layer number, $n_2$ varies non-monotonically with pump power, perhaps with slight oscillations. This, along with the power-dependent loss in **Figures 3(e)** and **(f),** can be attributed to power-sensitive photo-thermal changes of GO in the MRRs at high power, as noted previously [43, 69], as well as self-heating and thermal dissipation in the multilayer GO films. These effects could lead to a dynamic change in the GO material properties (e.g., linear refractive index, Kerr nonlinearity, linear/nonlinear absorption, and dispersion) with power, although in practice it is difficult to distinguish this from the effects of changes in loss and CE. In our FWM experiments, we slightly tuned the wavelength of the input CW light from blue to red around the resonances of the MRRs until it reached a steady thermal equilibrium state for FWM, and a temperature controller stage was used to maintain the steady state for long periods. In the steady state, the FWM CE did not show any time-dependent variation for fixed input pump power, indicating that any power-dependent change in the GO film's $n_2$ occurred very quickly and so any effects of this were not significant during the FWM experiments. We also note that the CW power in the MRRs (< 0.5 W) was not high enough to excite obvious satuable absorption of the GO films [27, 40]. Since it was difficult to accurately measure the slight change in film thickness and dispersion of GO with the pump power during FWM in the GO-coated MRRs, we neglected any change in these parameters. In principle, this approximation could lead to slight deviations in $n_2$, possibly explaining the non-monotonic relationship between $n_2$ and pump power. Despite this, the fit $n_2$ can still be regarded as a parameter reflecting the over-all FWM performance at different pump powers.

To compare the photo-thermal changes of the GO films during FWM, we also tested the performance of FWM in doped silica waveguides uniformly coated with 1 layer of GO or patterned with 50 layers of GO. The waveguides had the same geometry as those of the MRR, but with a longer length (~1.5 cm). This resulted in the generated idlers being above the noise floor. The GO coating length of the patterned waveguide (50 µm) was the same as that of the patterned MRR. The $n_2$ of GO films versus pump power obtained from FWM measurements using the GO-coated waveguides are plotted in **Figures 7(d)** and **(e)**, together with those obtained from FWM measurements using the GO-coated MRRs. As can be seen, $n_2$ shows a much greater variation with pump power in the MRRs than the waveguides, reflecting a more dramatic change of GO material properties in the MRRs. This is not surprising since the power is significantly higher in the MRRs.

**Figures 7(e)** and **(f)** present the transmission spectra of GO-coated MRRs before and after FWM (for a pump power of 22 dBm). No significant change was observed, and the measured CEs in **Figures 5(e) and (f)** were repeatable when we reinjected the CW pump and signal for FWM, indicating that the optically induced changes (e.g., loss, $n_2$) of the GO films in the MRRs were not permanent. Note that a variation in bandgap and material properties of GO can be obtained by changing GO's chemical structure and we previously showed [27, 37, 49] that the material properties of GO can also be permanently changed by femtosecond laser pulses with significantly higher power levels than those used here. This is distinct from the photo-thermal changes observed here.

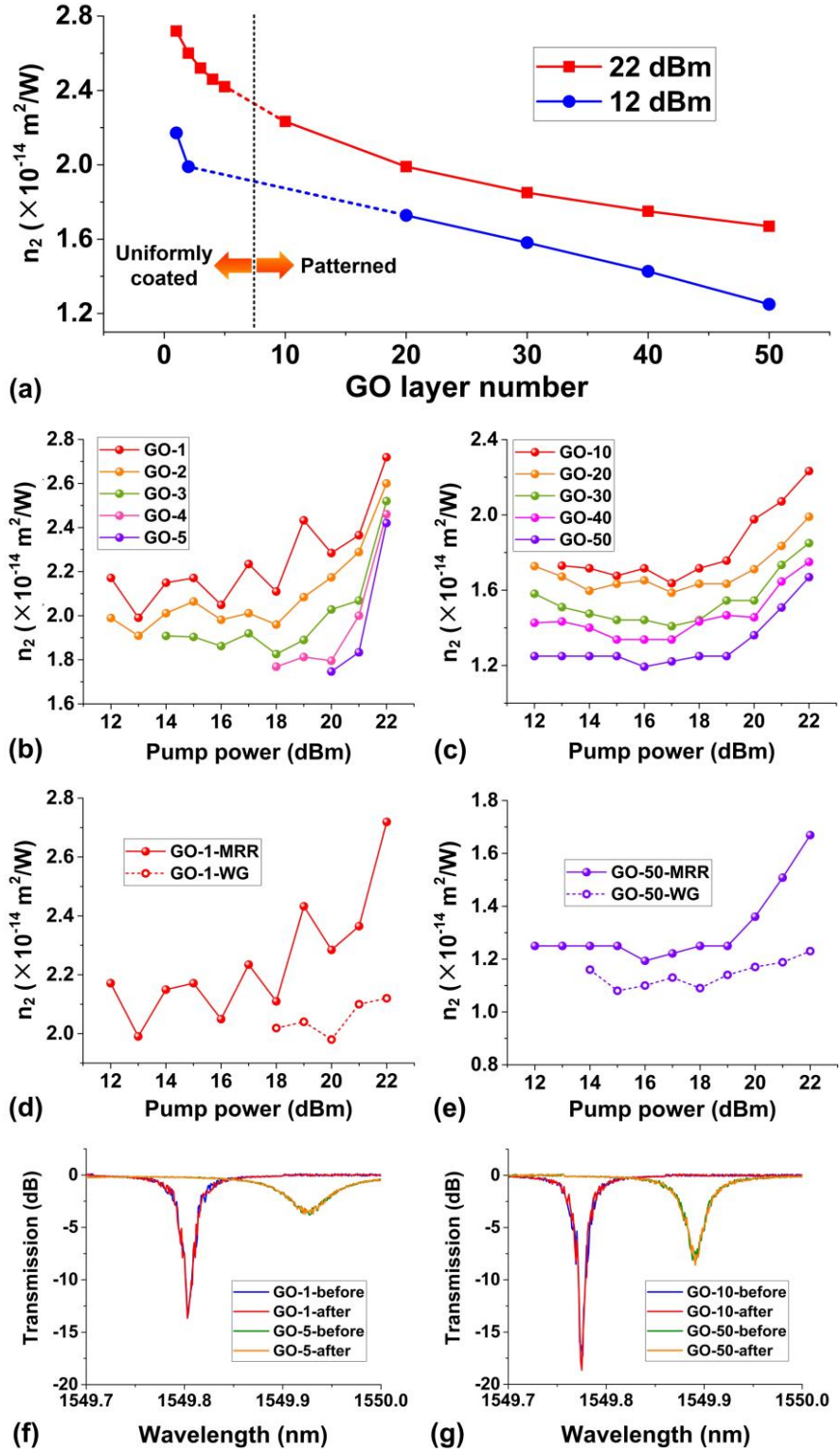

**Figure 7.** Analyses of the change of GO material properties during the FWM process. (a) $n_2$ of GO versus layer number at fixed pump powers of 12 dBm and 22 dBm obtained from the MRR FWM experiment. (b)−(c) $n_2$ of GO versus pump power for 1−5 layers and 10−50 layers of GO obtained from the MRR FWM experiment, respectively. (d)−(e) $n_2$ of GO versus pump power (1 layer of GO in (d) and 50 layers of GO in (e)) obtained from the MRR and waveguide (WG) FWM experiments. (f)−(g) Measured transmission spectra of GO-coated MRRs

(uniformly coated with 1 and 5 layers of GO in (f) and patterned with 10 and 50 layers of GO in (g)) before and after FWM with a pump power of 22 dBm.

Finally, FWM in MRRs depends on many factors in terms of material properties, including the third-order nonlinearity, the linear and nonlinear loss, dispersion, etc. For the GO-coated MRRs in our case, the physics of FWM are more complex due to the change of GO material properties with layer number and light power. Despite this, the layer-number and power-dependent material properties of the layered GO film yield many new device properties that are difficult to achieve for typical integrated photonic devices. We believe this could enable one to tailor the device performance well beyond simply enhancing the FWM CE as reported here.

## 5. Conclusion

We demonstrate enhanced FWM in MRRs integrated with layered GO films. CMOS-compatible doped silica MRRs with both uniformly coated and patterned GO films are fabricated based on a large-area, transfer-free, layer-by-layer GO coating method together with photolithography and lift-off processes, which yield precise control of the film thickness, placement, and coating length. We perform FWM measurements for MRRs uniformly coated with 1−5 layers of GO and patterned with 10−50 layers of GO, achieving up to ~7.6-dB and ~10.3-dB enhancement in the FWM CE for the MRRs uniformly coated with 1 layer of GO and patterned with 50 layers of GO, respectively. We also fit the measured CE to theory and obtain the change in the third-order nonlinearity of GO films with layer number and pump power. The high nonlinear performance of the GO-coated MRRs confirms the effectiveness of introducing 2D layered GO films into integrated photonic resonators to improve the performance of nonlinear optical processes.

## Acknowledgements

This work was supported by the Australian Research Council Discovery Projects Programs (No. DP150102972 and DP190103186), the Swinburne ECR-SUPRA program, the Industrial Transformation Training Centers scheme (Grant No. IC180100005), and the Beijing Natural Science Foundation (No. Z180007). The authors also acknowledge the Swinburne Nano Lab for the support in device fabrication and characterization. RM acknowledges support by the Natural Sciences and Engineering Research Council of Canada (NSERC) through the Strategic,

Discovery and Acceleration Grants Schemes, by the MESI PSR-SIIRI Initiative in Quebec, and by the Canada Research Chair Program. He also acknowledges additional support by the Professorship Program (grant 074-U 01) and by the 1000 Talents Sichuan Program in China.

Received: (will be filled in by the editorial staff)
Revised: (will be filled in by the editorial staff)
Published online: (will be filled in by the editorial staff)